\documentclass{jpsj3}
\usepackage{txfonts}
\usepackage{graphicx}
\usepackage{dcolumn}
\usepackage{bm}

\def\lesssim{\ \raise.3ex\hbox{$<$}\kern-0.8em\lower.7ex\hbox{$\sim$}\ }
\def\gesim{\ \raise.3ex\hbox{$>$}\kern-0.8em\lower.7ex\hbox{$\sim$}\ }

\title{Specific Heat and Effects of Uniaxial Anisotropy of a $p$-wave Pairing Interaction in a Strongly Interacting Ultracold Fermi Gas}

\author{Daisuke Inotani\thanks{dinotani@rk.phys.keio.ac.jp}, Pieter van Wyk, Yoji Ohashi}
\inst{Department of Physics, Keio University, 3-14-1 Hiyoshi, Kohoku-ku, Yokohama 223-8522, Japan} 

\abst{We investigate the specific heat $C_V$ at constant volume and effects of uniaxial anisotropy of a $p$-wave attractive interaction in the normal state of an ultracold Fermi gas. Within the framework of the strong-coupling theory developed by Nozi\`eres and Schmitt-Rink, we evaluate this thermodynamic quantity as a function of temperature, in the whole interaction regime. While the uniaxial anisotropy is not crucial for $C_V$ in the weak-coupling regime, $C_V$ is found to be sensitive to the uniaxial anisotropy in the strong-coupling regime. This originates from the population imbalance among $p_i$-wave molecules ($i=x,y,z$), indicating that the specific heat is a useful observable to see which kinds of $p$-wave molecules dominantly exist in the strong-coupling regime when the $p$-wave interaction has uniaxial anisotropy. Using this strong point, we classify the strong-coupling regime into some characteristic regions. Since a $p$-wave pairing interaction with uniaxial anisotropy has been discovered in a $^{40}$K Fermi gas, our results would be useful in considering strong-coupling properties of a $p$-wave interacting Fermi gas, when the interaction is uniaxially anisotropic. 
} 
\begin{document}
\maketitle
\par
\section{Introduction}
Feshbach resonance is one of the most crucial key phenomena in cold Fermi gas physics, because it enables us to tune the strength of a pairing interaction between Fermi atoms\cite{Chin,Timmermans}. In particular, $s$-wave and $p$-wave Feshbach resonances have been extensively discussed in $^{40}$K and $^6$Li Fermi gases. In the $s$-wave case, the so-called BCS (Bardeen-Cooper-Schrieffer)-BEC (Bose-Einstein condensation) crossover has been realized\cite{Regals,Zwierleins,Kinasts,Bartens}, where the character of a Fermi superfluid continuously changes from the weak-coupling BCS-type to the BEC of tightly bound molecules, with increasing the $s$-wave interaction strength by adjusting the threshold energy of a Feshbach resonance\cite{Eagles,NSR,Randeria,Engelbrecht,Haussmann,Leggett,OhashiGriffin,Zwerger,HuiHu}. In the normal state above the $s$-wave superfluid phase transition temperature $T_{\rm c}$, it has been discussed how a normal Fermi gas smoothly changes to a molecular Bose gas, as one passes through the BCS-BEC crossover region\cite{Perali,Tsuchiya1,Tsuchiya2,Pieter}. In the intermediate-coupling regime, various many-body phenomena have been studied, such as a pseudogap\cite{Perali,Tsuchiya1,Tsuchiya2,Jinphoto1,Jinphoto2,Kohl,Jinphoto3} and a spin-gap phenomenon\cite{Sanner,Tajima}. Although these have not been observed in the $p$-wave case, experimental techniques developed in the $s$-wave system would be also applicable to the $p$-wave case.
\par
In considering a $p$-wave Feshbach resonance, a key experimental report is the splitting of a $p$-wave Feshbach resonance observed in a $^{40}$K Fermi gas\cite{Ticknor}. To explain this phenomenon in a simple manner, we assume a $p$-wave Feshbach molecule with two atomic spins being polarized to be parallel to an external magnetic field in the $x$-direction. In this case, a magnetic dipole-dipole interaction gives different energy shifts between a Feshbach molecule with the total orbital angular momentum $L_x=0$ and a molecule with $L_x=\pm 1$. Since the dipole-dipole interaction is more repulsive in the latter case than the former, the threshold energy of a $p$-wave Feshbach resonance becomes larger in the latter than the former, leading to the observed splitting of a $p$-wave Feshbach resonance\cite{Ticknor}.
\par
When one uses this $p$-wave Feshbach resonance to produce a $p$-wave pairing interaction $V_p({\bm p},{\bm p}')$\cite{Ticknor,Regal,Regal2,Zhang,Schunck,Gunter,Gaebler2,Inaba,Fuchs,Mukaiyama,Maier}, it has a uniaxial anisotropy as,
\begin{equation}
V_p({\bm p},{\bm p}')=-\left( U_xp_xp_x'+U_yp_yp_y'+U_zp_zp_z' \right) ~~(U_x>U_y=U_z),
\label{eq01}
\end{equation}
where the inequality $U_x>U_y=U_z$ is because one first meets the non-degenerate $L_x=0$-channel with decreasing an external magnetic field to approach the $p$-wave Feshbach resonance. 
\par
Although a $p$-wave interaction is anisotropic by nature, this additional uniaxial anisotropy is expected to further enrich the superfluid phase diagram of this system\cite{Gurarie,Gurarie2,Inotani3}. For example, while the $p_x+ip_y$-wave superfluid phase is only possible when $U_x=U_y=U_z$, multi-superfluid phase has theoretically been predicted when $U_x>U_y=U_z$\cite{Gurarie,Gurarie2,Inotani3}. According to this prediction, $p_x$-wave superfluid phase transition first occurs at $T_{\rm c}$, which is followed by the second $p_x+ip_y$-wave phase transition at $T_{\rm c}'<T_{\rm c}$. In the $p_x$-wave superfluid phase near $T_{\rm c}'$, the possibility of pseudogap phenomena caused by $p_x+ip_y$-wave pairing fluctuations has also been discussed\cite{Inotani3}. While such a multi-superfluid phase is absent in the ordinary $s$-wave Fermi superfluid, it has been discussed in superfluid $^3$He\cite{He3}, as well as unconventional superconductors, such as UPt$_3$\cite{Bruls,Sigrist}. Together with the tunability of a $p$-wave pairing interaction\cite{Chin}, an ultracold Fermi gas with a uniaxially anisotropic $p$-wave interaction is expected to be a useful quantum simulator for the study of multi-superfluid phase in a systematic manner.
\par
Since the uniaxial anisotropy of a $p$-wave interaction has so far been mainly focused on the viewpoint of superfluid physics\cite{Gurarie,Gurarie2,Inotani3}, it has not been studied in detail how the anisotropy influences normal state properties. The purpose of this paper is just to clarify this normal-state problem. Extending the BCS (Bardeen-Cooper-Schrieffer)-BEC (Bose-Einstein condensation) crossover theory developed by Nozi\`eres and Schmitt-Rink (NSR)\cite{NSR} to the case of a uniaxially anisotropic $p$-wave interaction, we show that this anisotropy remarkably affects the normal-state behavior of the specific heat $C_V$ at constant volume in the strong-coupling regime. As the background physics of this, we point out the importance of population imbalance among three kinds of $p_i$-wave bound molecules ($i=x,y,z$) which is caused by the uniaxial anisotropy. We also show that the temperature dependence of this thermodynamic quantity involves useful information about which kinds of $p$-wave pairs are dominantly formed in the phase diagram of a $p$-wave interacting Fermi gas.
\par
Before ending this section, we comment on the current stage of research on $p$-wave interacting Fermi gases. The splitting of a $p$-wave Feshbach resonance has been observed in a one-component $^{40}$K Fermi gas\cite{Ticknor}. In this one-component case, the contact-type $s$-wave interaction is forbidden by the Pauli's exclusion principle, so that a $p$-wave interaction can be the leading interaction. In addition, the so-called dipolar loss is suppressed in the one-component case\cite{Chevy}, so that a one-component Fermi gas is usually used to explore a $p$-wave superfluid phase transition (although no one has achieved $T_{\rm c}$). Thus, in this paper, we deal with a one-component Fermi gas with a uniaxially anisotropic $p$-wave pairing interaction. For the specific heat $C_V$ at constant volume, it has recently become observable in cold Fermi gas physics\cite{Ku}. Although this thermodynamic quantity has only been measured in an $s$-wave interacting Fermi gas, the same technique would also be applicable to the $p$-wave case. Theoretically, Ref. \cite{Pieter} has recently examined the specific heat $C_V$ in an $s$-wave interacting Fermi gas. References \cite{Inotani4,Inotani5} extended this work to the $p$-wave case in the absence of uniaxial anisotropy.
\par
This paper is organized as follows. In Sec. II, we explain our formulation. In Sec. III, we examine effects of uniaxial anisotropy of a $p$-wave pairing interaction on the specific heat $C_V$ at $T_{\rm c}$. We extend our discussions to the region above $T_{\rm c}$ in Sec. IV. Here, we examine how the detailed temperature dependence of $C_V$ reflects the population imbalance of three kinds of $p$-wave molecules caused by the uniaxial anisotropy of the $p$-wave interaction. Throughout this paper, we take $\hbar=k_{\rm B}=1$, and the system volume $V$ is taken to be unity, for simplicity.
\par
\par
\section{Formulation}
\par
We consider a model one-component Fermi gas with a uniaxially anisotropic $p$-wave interaction $V_p({\bm p},{\bm p}')$, described by the BCS-type Hamiltonian\cite{Inotani3,Ohashi,Ho,Botelho,Iskin,Iskin2,Cheng,Inotani,Inotani2},
\begin{equation}
H=\sum_{\bm p} \xi_{\bm p}c_{\bm p}^{\dagger}c_{\bm p}
+\frac{1}{2}\sum_{{\bm p},{\bm p}',{\bm q}} 
V_p({\bm p},{\bm p}')
c_{{\bm p}+{\bm q}/2}^\dagger c_{-{\bm p}+{\bm q}/2}^\dagger
c_{-{\bm p}'+{\bm q}/2}c_{{\bm p}'+{\bm q}/2},
\label{eq1}
\end{equation}
where $c^\dagger_{\bm p}$ is the creation operator of a Fermi atom with the kinetic energy $\xi_{\bm p}=\varepsilon_{\bm p}-\mu={\bm p}^2/(2m)-\mu$, measured from the Fermi chemical potential $\mu$ (where $m$ is an atomic mass). The $p$-wave attractive interaction $V_p({\bm p},{\bm p}')$ has the separable form\cite{Ho,Inotani3},
\begin{equation}
V_p({\bm p},{\bm p}')=-\sum_{i=x,y,z} 
\gamma^i_{\bm p} 
U_i \gamma^i_{\bm {p}'}.
\label{eq2}
\end{equation}
Here, $-U_i$ ($<0$) is a coupling constant in the $p_i$-wave Cooper channel ($i=x,y,z$), which is assumed to be tunable by a $p$-wave Feshbach resonance. We model the uniaxial anisotropy observed in a $^{40}$K Fermi gas\cite{Ticknor} by taking $U_x \ge U_y=U_z$, where $x$-axis is taken to be parallel to an external magnetic field to experimentally tune a $p$-wave Feshbach resonance. In Eq. (\ref{eq2}), 
\begin{equation}
\gamma^i_{\bm p}=p_i F_{\rm c}({\bm p})  
\label{eq3}
\end{equation}
is a $p$-wave basis function, where $F_{\rm c}({\bm p})=1/[1+(p/p_{\rm c})^6]$ is a cutoff function, to eliminate the ultraviolet divergence coming from the $p$-wave interaction $V_p({\bm p},{\bm p}')$ in Eq. (\ref{eq2}). We briefly note that the detailed momentum dependence of the cutoff function $F_{\rm c}({\bm p})$ actually does not affect normal-state quantities, as far as we take the cutoff momentum $p_{\rm c}$ to be much larger than the Fermi momentum $k_{\rm F}$\cite{Inotani3}. As usual, we relate the bare coupling constants $U_i$ ($i=x,y,z$), as well as the cutoff momentum $p_{\rm c}$, to the observable $p$-wave scattering volumes $v_i$ and the inverse $p$-wave effective range $k_0$, as
\begin{eqnarray}
\left\{
\begin{array}{l}
\displaystyle
{4\pi v_i \over m}=-
{U_i \over 3}
{1 \over \displaystyle 1-{U_i \over 3}\sum_{\bm p}
{{\bm p}^2 \over 2\varepsilon_{\bm p}}F_{\rm c}^2({\bm p})},
\\
\displaystyle
k_0=-{4\pi \over m^2}
\sum_{\bm p}
{{\bm p}^2 \over 2\varepsilon_{\bm p}^2}F_{\rm c}^2({\bm p}).
\end{array}
\right.
\label{eq5}
\end{eqnarray}
Although $k_0$ may be different among the three $p_i$-wave Cooper channel, we have ignored this channel dependence in Eq. (\ref{eq5}). Following the experiment on a $^{40}$K Fermi gas\cite{Ticknor}, we take $k_0=-30k_{\rm F}$. The $p_i$-wave interaction strength can then be measured in terms of $(k_{\rm F}^3v_i)^{-1}$. $(k_{\rm F}^3v_i)^{-1}\lesssim 0$ and $(k_{\rm F}^3v_i)^{-1}\gesim 0$ characterize the weak-couping side and the strong-coupling side, respectively. To describe the uniaxial anisotropy, we also introduce the anisotropy parameter, 
\begin{equation}
\delta v^{-1}\equiv v^{-1}_x-v^{-1}_y=v^{-1}_x-v^{-1}_z~(>0).
\label{eq5b}
\end{equation} 
The interaction strength is then completely determined by $(k_{\rm F}^3v_x)^{-1}$, and $(k_{\rm F}^3\delta v)^{-1}$.
\par
\begin{figure}
\centerline{\includegraphics[width=8cm]{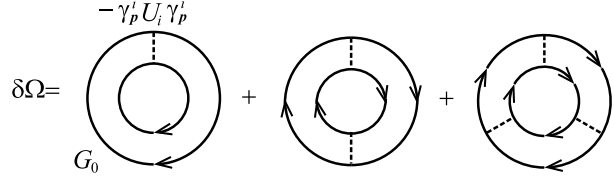}}
\caption{Feynman diagrams describing fluctuation corrections $\delta\Omega$ to the thermodynamic potential in the NSR theory. The solid lines and dashed lines describe the bare single-particle thermal Green's function, $G_0({\bm p}, i\omega_n)=[i\omega_n-\xi_p]^{-1}$ (where $\omega_n$ is the fermion Matsubara frequency), and the $p$-wave interaction $V_p({\bm p},{\bm p}')$, respectively.}
\label{fig1}
\end{figure}
\par
We treat strong-coupling effects of the $p$-wave pairing interaction within the NSR theory\cite{NSR}. In this scheme, strong-coupling corrections $\delta\Omega$ to the thermodynamic potential $\Omega$ are diagrammatically given as Fig. \ref{fig1}. Summing up these diagrams, we obtain
\begin{equation}
\delta\Omega=T\sum_{{\bm q},\nu_n}
{\rm Tr}\ln \left[1-{\hat U}_p{\hat \Pi}({\bm q}, i\nu_n)\right].
\label{eq8}
\end{equation}
Here, ${\hat U}_P={\rm diag}[U_x,U_y,U_z]$, and ${\hat \Pi}=\{\Pi_{i,j} \}$ ($i,j=x,y,z$) is the $3\times 3$-matrix lowest-order pair-correlation function, where
\begin{eqnarray}
\Pi_{i,j}({\bm q},i\nu_n)= 
\frac{1}{2}\sum_{\bm {p}} 
\gamma^i_{\bm{p}} 
\frac{\tanh \left( \frac{\xi_{\bm{p} + \frac{\bm q}{2}}}{2T} \right)
+\tanh \left( \frac{\xi_{-\bm{p} + \frac{\bm q}{2}}}{2T} \right)}
{i\nu_n-\xi_{\bm{p} + \frac{\bm q}{2}}-\xi_{-\bm{p} + \frac{\bm q}{2}}}
\gamma^j_{\bm{p}}.
\label{eq10}
\end{eqnarray}
The total thermodynamic potential $\Omega$ is then given by
\begin{equation}
\Omega=\Omega_0+\delta\Omega,
\label{eq6}
\end{equation}
where 
\begin{equation}
\Omega_0=T \sum_{{\bm p}} \ln \left[1+e^{-\xi_{\bm p}/T} \right]
\label{eq7}
\end{equation}
is the thermodynamic potential in a free Fermi gas. 
\par
The specific heat $C_V(T)$ at constant volume can be calculated from the internal energy $E$ as,
\begin{eqnarray}
C_V(T)=\left(
\frac{ \partial E}{\partial T}
\right)_{V,N}.
\label{eq14}
\end{eqnarray}
In the NSR formalism, the internal energy $E$ is conveniently obtained from the thermodynamic potential $\Omega$ in Eq. (\ref{eq6}) by using the Legendre transformation as,
\begin{eqnarray}
E&=&
\Omega
-T\left( \frac{\partial \Omega }{\partial T} \right)_{\mu}
-\mu\left( \frac{\partial \Omega }{\partial \mu} \right)_{T}
\nonumber
\\
& &
\hskip-12mm
=
\sum_{{\bm p}}\varepsilon_{\bm p} f(\xi_{\bm p})
\nonumber
\\
& &
\hskip-12mm
-T\sum_{{\bm q},\nu_n}
{\rm Tr}
\left[
{\hat \Gamma}({\bm q},i\nu_n)
\left[
T\left(\frac{\partial {\hat \Pi}({\bm q},i\nu_n)}{\partial T}\right)_\mu
+\mu\left(\frac{\partial {\hat \Pi}({\bm q},i\nu_n)}{\partial \mu}\right)_T
\right]
\right],
\nonumber
\\
\label{eq15}
\end{eqnarray}
where $f(\xi_{\bm p})$ is the Fermi distribution function, and 
\begin{equation}
{\hat \Gamma}({\bm q},i\nu_n)=
{-{\hat U}_p \over 1-{\hat U}_p{\hat \Pi}({\bm q},i\nu_n)}
\label{eq.Gamma}
\end{equation}
is the $3\times 3$-matrix particle-particle scattering matrix in the $T$-matrix approximation.
\par
In calculating the specific heat $C_V$ above $T_{\rm c}$, we need to evaluate the Fermi chemical potential $\mu(T)$, which is achieved by considering the equation for the total number $N$ of Fermi atoms involving effects of $p$-wave pairing fluctuations, given by
\begin{eqnarray}
N&=&-\left( \frac{\partial \Omega}{\partial \mu} \right)_{T}
\nonumber
\\
&=&
\sum_{\bm p}f(\xi_{\bm p})
+
T\sum_{{\bm q},\nu_n}
{\rm Tr}
\left[\Gamma({\bm q},i\nu_n)
\left(\frac{\partial {\hat \Pi}({\bm q},i\nu_n)}{\partial \mu} \right)_T
\right].
\nonumber
\\
\label{eq11}
\end{eqnarray}
\par
As well known in the ordinary NSR theory, the last term in the second line in Eq. (\ref{eq11}) is reduced to twice the number $N_{\rm B}$ of tightly bound molecules in the strong-coupling regime ($(k_{\rm F}^3v_x)^{-1}\gg 1$). In the $p$-wave case, it is convenient write this term as $2N_{\rm B}=2\sum_{i=x,y,z}N_{\rm B}^i$, where
\begin{equation}
N_{\rm B}^i=
{T \over 2}\sum_{{\bm q},\nu_n}
\left[
{\hat \Gamma}({\bm q},i\nu_n)
\left(\frac{\partial {\hat \Pi}({\bm q},i\nu_n)}{\partial \mu} \right)_T
\right]_{i,i}
\label{eq11b}
\end{equation}
is the number of $p_i$-wave molecules in the strong-coupling regime. 
\par
Since we are dealing with the uniaxially anisotropic $p$-wave interaction ($U_x>U_y=U_z$), the superfluid instability first occurs in the $p_x$-wave Cooper channel. Thus, the equation for $T_{\rm c}$ is obtained from the Thouless criterion in this channel. That is, the superfluid instability occurs, when the particle-particle scattering matrix $\Gamma_{x,x}({\bm q},i\nu_n)$ in Eq. (\ref{eq.Gamma}) has a pole at ${\bm q}=\nu_n=0$, which gives, 
\begin{eqnarray}
1=U_x \sum_{\bm p} \frac{({\gamma^x_{\bm p}})^2 }{2\xi_p} \tanh \left( \frac{\xi_p}{2T} \right).
\label{eq13}
\end{eqnarray}
\par
For a given interaction strength, we numerically solve the $T_{\rm c}$-equation (\ref{eq13}), together with the number equation (\ref{eq11}), to determined $T_{\rm c}$ and $\mu(T_{\rm c})$ self-consistently. In the normal state above $T_{\rm c}$, we only solve the number equation (\ref{eq11}) to determine $\mu(T)$. We then numerically execute the derivative in Eq. (\ref{eq14}) by calculating the internal energy $E$ in Eq. (\ref{eq15}) at slightly different two temperatures\cite{Pieter}.
\par
\begin{figure}
\centerline{\includegraphics[width=8cm]{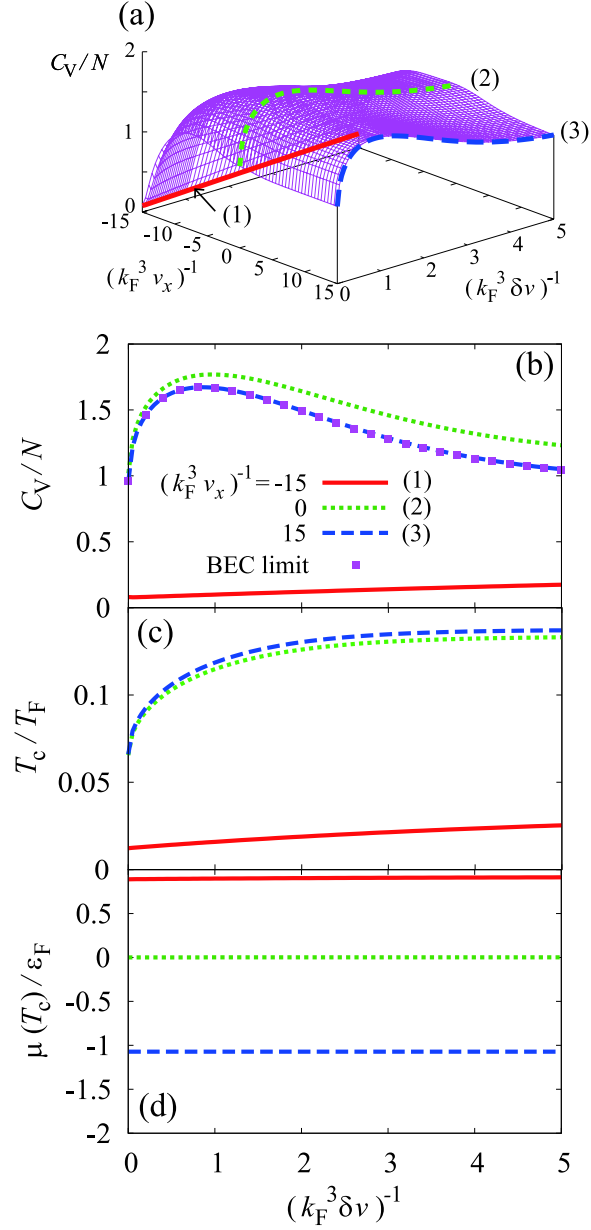}}
\caption{(Color online) (a) Calculated specific heat $C_V$ at $T_{\rm c}$, as functions of the $p_x$-wave interaction strength $(k_{\rm F}^3v_x)^{-1}$ and the anisotropy parameter $(k_{\rm F}^3\delta v)^{-1}$. For clarify, the results at the three typical interaction strengths (1)-(3) are re-drawn in panel (b). The solid squares in this panel show $C_V$ in the BEC limit obtained from Eq. ($\ref{eq17}$). (c) $T_{\rm c}$ as a function of $(k_{\rm F}^3\delta v)^{-1}$. (d) Fermi chemical potential $\mu$ at $T_{\rm c}$. $T_{\rm F}$ and $\varepsilon_{\rm F}$ are the Fermi temperature and the Fermi energy, respectively.}
\label{fig2}
\end{figure}
\par
\begin{figure}
\centerline{\includegraphics[width=7cm]{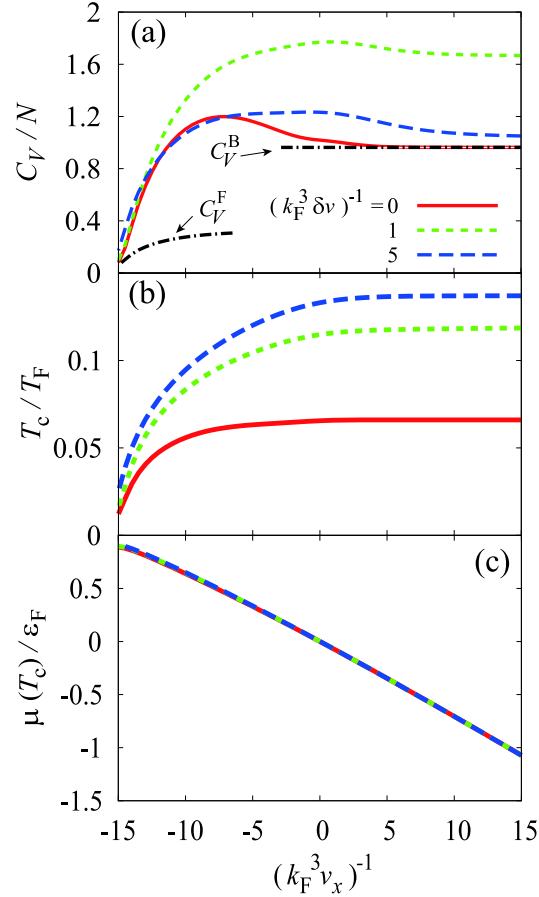}}
\caption{(Color online) (a) Calculated specific heat $C_V$ at $T_{\rm c}$, as a function of the $p$-wave interaction strength $(k_{\rm F}^3v_x)^{-1}$. $C_V^{\rm F}$ is the specific heat in a free Fermi gas at $T_{\rm c}$. $C_V^{\rm B}$ is the specific heat at $T_{\rm BEC}$ of an ideal Bose gas, consisting of three kinds of $N/6$ $p$-wave molecules. (b) Superfluid phase transition temperature $T_{\rm c}$.  (c) Fermi chemical potential $\mu(T_{\rm c})$. Note that the three cases ($(k_{\rm F}^3\delta v)^{-1}=0,~1,~5$) give almost the same value of $\mu$.
}
\label{fig3}
\end{figure}
\par
\section{Specific heat $C_V$ and effects of uniaxial anisotropy of $p$-wave interaction at $T_{\rm c}$}
\par
Figure \ref{fig2}(a) shows the specific heat $C_V$ at $T_{\rm c}$, as functions of the interaction strength $v_x^{-1}$ and the anisotropy parameter $\delta v^{-1}$. In the weak-coupling regime ($(k_{\rm F}^3v_x)^{-1}\lesssim -10$), $C_V(T_{\rm c})$ is not so sensitive to the anisotropy parameter $\delta v^{-1}$. As the interaction strength $v_x^{-1}$ increases, however, $C_V$ as a function of $\delta v^{-1}$ exhibits a hump structure. To see this more clearly, we re-plot the typical three cases in Fig. \ref{fig2}(b), where one sees that the hump is located at $(k_{\rm F}\delta v)^{-1}\sim 1$, when $(k_{\rm F}^3v_x)^{-1}\gesim 0$.
\par
To understand the $\delta v^{-1}$-dependence of the specific heat seen in the cases of (2) and (3) shown in Fig. \ref{fig2}(b), it is convenient to consider the extreme strong-coupling regime where the Fermi chemical potential is negative and satisfies $|\mu(T_{\rm c})|\gg T$. In this extreme case, the particle-particle scattering matrix ${\hat \Gamma}({\bm q},i\nu_n)$ in Eq. (\ref{eq.Gamma}), which physically describes $p$-wave pairing fluctuations, is reduced to the molecular Bose Green's function as,
\begin{eqnarray}
\Gamma_{i,j}({\bm q},i\nu_n) \simeq \frac{24}{m^2|k_0|} \frac{1}{i\nu_n-\xi^i_{\rm B}({\bm q})}\delta_{i,j},
\label{eq16}
\end{eqnarray}
where 
\begin{equation}
\xi^i_{\rm B} ({\bm q})={{\bm q}^2 \over 2M_{\rm B}}-\mu_{\rm B}^i
\label{eq16b}
\end{equation}
is the kinetic energy of a Bose molecule in the $p_i$-wave Cooper channel ($i=x,y,z$). $M_{\rm B}=2m$ is a molecular mass, $\mu_{\rm B}^i=2\mu-E_{\rm bind}^i$ is a molecular chemical potential, where
\begin{equation}
E_{\rm {bind}}^i =-{2 \over m|k_0|v_i}
\label{eq16c}
\end{equation}
is the binding energy of a $p_i$-wave two-body bound state\cite{Ho}. Noting that in the strong-coupling limit $\mu \simeq E_{\rm {bind}}^x/2$, the specific heat in this extreme case is then evaluated as,
\begin{eqnarray}
C_V=\sum_{i=x,y,z}\frac{\partial}{\partial T} \left[
 \sum_{\bm q} \frac{q^2}{2M_{\rm B}}
n_{\rm B}(\xi_{\rm B}^i({\bm q}))+E_{\rm {bind}}^i N_{\rm B}^i
\right],
\label{eq17}
\end{eqnarray}
where $n_{\rm B}(\xi_{\rm B}^i({\bm q}))$ is the Bose distribution function, and $N_{\rm B}^i$ is the number of tightly bound molecules in the $p_i$-wave Cooper channel in Eq. (\ref{eq11b}). As shown in Fig. \ref{fig2}(b), Eq. (\ref{eq17}) well reproduces $C_V$ when $(k_{\rm F}^3v_x)^{-1}=15$.
\par
In Eq. (\ref{eq17}), the first term in the brackets is the ordinary expression for the specific heat in an ideal Bose gas. Noting that (1) $T_{\rm c}$ monotonically increases with increasing the magnitude of the anisotropy parameter $\delta v^{-1}$ (see Fig. \ref{fig2}(c)), and (2) the Fermi chemical potential $\mu(T_{\rm T})$ is almost independent of $\delta v^{-1}$ (see Fig. \ref{fig2}(d)), one finds that this term cannot explain the hump seen in Fig. \ref{fig2}(b). 
\par
For the second term in the brackets in Eq. (\ref{eq17}), since all the Fermi atoms form two-body bound molecules in the strong-coupling limit, one finds $N_{\rm B}^x+N_{\rm B}^y+N_{\rm B}^z=N_{\rm B}^x+2N_{\rm B}^y=N/2$. In this case, the contribution from this term ($\equiv C_V^{(2)}$) is evaluated as,
\begin{eqnarray}
C_V^{(2)}&\simeq& 
{\partial \over \partial T}
\left[
2\bigl[E_{\rm bind}^y-E_{\rm bind}^x\bigr]N_{\rm B}^y+E_{\rm bind}^xN
\right]={4 \over m|k_0|\delta v}
{\partial N_{\rm B}^y \over \partial T}.
\nonumber
\\
\label{eq17b}
\end{eqnarray}
When $U_x>U_y=U_z$, the binding energy $E_{\rm bind}^x$ in the $p_x$-wave channel is lower than $E_{\rm bind}^y$ ($=E_{\rm bind}^z$). Thus, while $p_x$-wave molecules become dominant ($N_{\rm B}^x\simeq N/2$) when $T\ll\Delta E\equiv E_{\rm bind}^y-E_{\rm bind}^x=2/(m|k_0|\delta v)$, the increase of $N_{\rm B}^y$ around $T=2/(m|k_0|\delta v)$ is expected to enhance $C_V^{(2)}$ in Eq. (\ref{eq17b}). Indeed, around the peak position ($(k_{\rm F}^3\delta v)^{-1}\simeq 1$) in Fig. \ref{fig2}(b), one finds,
\begin{equation}
{2 \over m|k_0|\delta v}\simeq 0.13T_{\rm F},
\label{eq17c}
\end{equation}
which is comparable to $T_{\rm c}=0.12T_{\rm F}$ at $(k_{\rm F}^3\delta v)^{-1}=1$ (see Fig. \ref{fig2}(c)). Thus, the hump structure seen in Fig. \ref{fig2}(b) is due to thermal excitations of $p_x$-wave molecules into $p_y$- and $p_z$-wave molecular states.
\par
We briefly note that $\partial N_{\rm B}^y/\partial T=0$ in the absence of uniaxial anisotropy when $T\ll E_{\rm bind}^i$. In this case, $C_V$ is dominated by the first term in the bracket in Eq. (\ref{eq17}), so that the low temperature behavior of $C_V$ is essentially the same as that in an ideal Bose gas of $p$-wave molecules.
\par
We also note that, deep inside the strong-coupling regime ($(k_{\rm F}^3v_x)^{-1}\gg 1$), most Fermi atoms form $p_x$-wave molecules when $E_{\rm bind}^{x}\ll E_{\rm bind}^{i=y,z}$, or equivalently $(k_{\rm F}^3\delta v)^{-1}\gg 1$. In this extreme case, $T_{\rm c}$ is simply obtained as the BEC transition temperature $T_{\rm BEC}(N/2)$ in an ideal Bose gas with $N/2$ $p_x$-wave molecules, given by\cite{Inotani3}
\begin{equation}
T_{\rm BEC}(N/2)={2\pi \over \zeta(3/2)M_{\rm B}}
\left(
{N \over 2}
\right)^{2/3}=0.137T_{\rm F},
\label{eq17d}
\end{equation}
where $\zeta(3/2)=2.612$ is the zeta function. On the other hand, when $U_x=U_y=U_z$, all the three $p_i$-wave molecular states ($i=x,y,z$) are equally populated, so that $T_{\rm c}$ in this limit equals $T_{\rm BEC}(N/6)$ of an ideal Bose gas with $N/6$ molecules, given by\cite{Inotani3,Ohashi,Ho}
\begin{equation}
T_{\rm BEC}(N/6)={2\pi \over \zeta(3/2)M_{\rm B}}
\left(
{N \over 6}
\right)^{2/3}=0.066T_{\rm F}.
\label{eq17e}
\end{equation}
Indeed, $T_{\rm c}$ in Fig. \ref{fig2}(c) continuously changes from $T_{\rm BEC}(N/6)$ to $T_{\rm BEC}(N/2)$, with increasing the anisotropy parameter $\delta v^{-1}$, when $(k_{\rm F}^3v_x)^{-1}=15$.
\par
Figure \ref{fig3}(a) shows the specific heat $C_V$ at $T_{\rm c}$, as a function of the $p$-wave interaction strength. When $\delta v^{-1}=0$, $C_V(T_{\rm c})$ gradually becomes larger than the free Fermi gas result $C_V^{\rm F}$ with increasing the interaction strength from the weak-coupling regime, to exhibit a peak around $(k_{\rm F}^3v_x)^{-1}=-7$. As discussed in Ref.\cite{Inotani5}, this peak structure originates from the enhancement of $p$-wave pairing fluctuations near $T_{\rm c}$. In the strong-coupling regime ($(k_{\rm F}^3v_x)^{-1}\gesim 5$) where $T_{\rm c}$ is almost constant and $\mu$ is negative (see Figs. \ref{fig3}(b) and (c)), $C_V(T_{\rm c})$ in the absence of uniaxial anisotropy approaches the specific heat $C_V^{\rm B}$ in an ideal Bose gas with three kinds of $N/6$ $p$-wave molecules at the BEC phase transition temperature $T_{\rm BEC}(N/6)$ in Eq. (\ref{eq17e}), given by
\begin{equation}
C_V^{\rm B}(T_{\rm BEC})=3{15 \over 4}
{\zeta(5/2) \over \zeta(3/2)}
\left({N \over 6}\right)
=0.963N,
\label{eq18}
\end{equation}
where $\zeta(5/2)=1.341$. The factor ``3" reflects the existence of three kinds of $p_i$-wave molecules ($i=x,y,z$).
\par
When $\delta v^{-1}>0$, Fig. \ref{fig3}(a) shows that the specific heat $C_V(T_{\rm c})$ still has a similar interaction dependence to the case of $\delta v^{-1}=0$ in the weak-coupling regime ($(k_{\rm F}^3 v_x)^{-1}\lesssim -7$). When $(k_{\rm F}^3\delta v)^{-1}=5$, Fig. \ref{fig3}(a) also shows that, apart from the detailed peak structure, $C_V(T_{\rm c})$ also looks approaching $C_V^{\rm B}(T_{\rm BEC})$ in Eq. (\ref{eq18}). In this highly anisotropic case, because of $E_{\rm bind}^x\ll E_{\rm bind}^{i=y,z}$, most Fermi atoms form $p_x$-wave molecules in the strong-coupling regime. Indeed, as seen in Figs. \ref{fig3}(b) and (c), $T_{\rm c}$ approaches $T_{\rm BEC}(N/2)$ in Eq. (\ref{eq17d}) when $\mu<0$. As a result, $C_V(T_{\rm c})$ in the strong-coupling regime is reduced to the specific heat in an ideal Bose gas with $N/2$ $p_x$-wave molecules, which is given by Eq. (\ref{eq18}) where the factor ``3" is absent and $N/6$ is replaced by $N/2$. The resulting specific heat has the same value as the ``three-component" case in Eq. (\ref{eq18}), which explains the strong-coupling behavior of $C_V(T_{\rm c})$ when $(k_{\rm F}^3\delta v)^{-1}=5$ in Fig. \ref{fig3}(a).
\par
When $(k_{\rm F}^3 \delta v)^{-1}=1$, one sees in Fig. \ref{fig3}(a) that the specific heat at $T_{\rm c}$ becomes larger than the other two cases shown in this figure, and it does not approach $C_V^{\rm B}(T_{\rm BEC})$ in Eq. (\ref{eq18}) even in the strong-coupling regime. In this case, as mentioned previously, since the value, $T_{\rm c}\simeq 0.12T_{\rm F}$, in the strong-coupling regime is comparable to the energy difference $\Delta E= E_{\rm bind}^{i=y,z}-E_{\rm bind}^x=2/(m|k_0|\delta v)$, $C_V^{(2)}$ in Eq. (\ref{eq17b}) enhances the specific heat $C_V(T_{\rm c})$. In addition, $\Delta E$ remains finite in the strong-coupling limit, so that the contribution $C_V^{(2)}$ continues to exist even in this limit, leading to the large $C_V(T_{\rm c})$ in the strong-coupling regime, compared to the cases of $(k_{\rm F}\delta v)^{-1}=0$ and $5$ in Fig. \ref{fig3}(a).
\par
\begin{figure}
\centerline{\includegraphics[width=8cm]{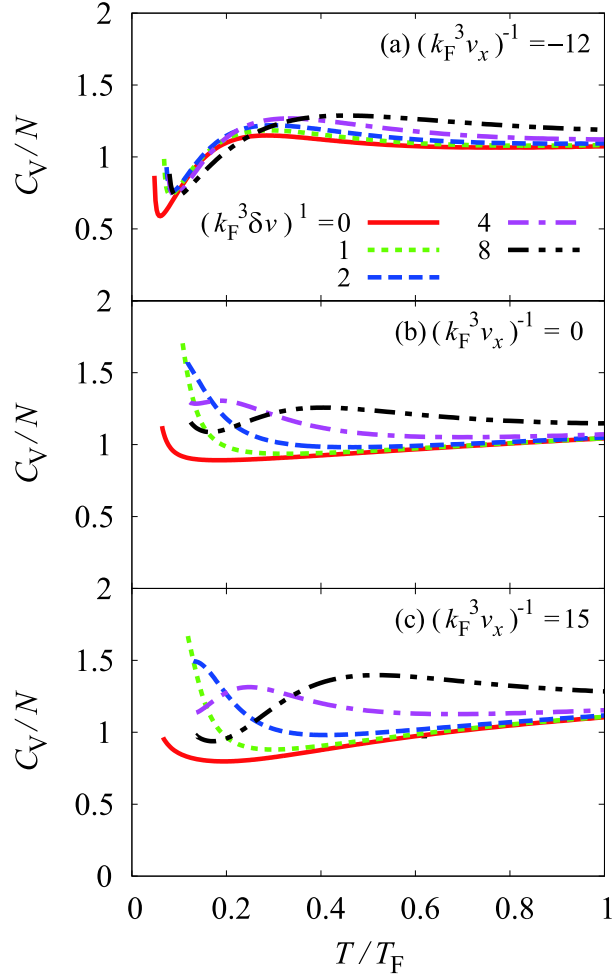}}
\caption{(Color online) Calculated specific heat $C_V(T)$ above $T_{\rm c}$. Three panels show typical examples in (a) the weak-coupling regime, (b) the intermediate-coupling regime, and (c) the strong-coupling regime.}
\label{fig4}
\end{figure}
\par
\section{Temperature dependence of the specific heat $C_V$ in a uniaxially anisotropic $p$-wave interacting Fermi gas}
\par
We now consider the specific heat $C_V(T)$ above $T_{\rm c}$. In Fig. \ref{fig4}(a), one finds that in the weak-coupling regime ($(k_{\rm F}^3 v_x)^{-1}=-12$), the uniaxial anisotropy of the $p$-wave interaction is not so crucial for $C_V(T)$ above $T_{\rm c}$, as expected from the previous discussions at $T_{\rm c}$.  That is, irrespective of the values of the anisotropy parameter $\delta v^{-1}$, one sees a dip structure near $T_{\rm c}$, as well as a hump at $0.2\lesssim T/T_{\rm F}\lesssim 0.4$. As discussed in Refs.\cite{Inotani4,Inotani5}, these structures originate from strong $p$-wave pairing fluctuations near $T_{\rm c}$, and anomalous particle-particle scatterings into $p$-wave molecular states, respectively.
\par
On the other hand, with increasing the interaction strength, effects of the uniaxial anisotropy are seen in Figs. \ref{fig4}(b) and (c). In these cases, the enhancement of the specific heat $C_V(T)$ near $T_{\rm c}$ becomes remarkable when $(k_{\rm F}\delta v)^{-1}=1$, but again $C_V(T\simeq T_{\rm c})$ becomes small with further increasing the anisotropy parameter $\delta v^{-1}$. Instead, a hump structure appears when $(k_{\rm F}\delta v)^{-1}\gesim 4$, the position of which  shifts to higher temperatures with increasing $\delta v^{-1}$. In addition, a dip structure revives near $T_{\rm c}$, when $(k_{\rm F}\delta v)^{-1}\gesim 8$. 
\par
\begin{figure}
\centerline{\includegraphics[width=8cm]{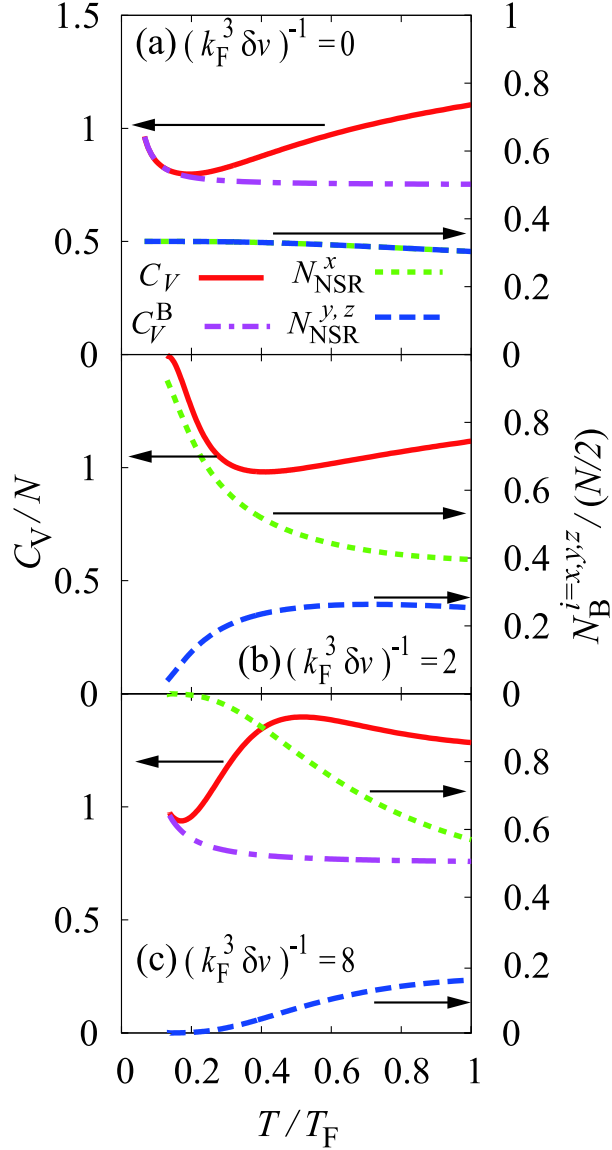}}
\caption{(Color online) Relation between the temperature dependence of $C_V(T)$ and the number $N_{\rm B}^{i=x,y,z}$ of $p_i$-wave molecular bosons in the strong-coupling regime ($(k_{\rm F}^3v_x)^{-1}=15$). $C_V^{\rm B}(T)$ in panel (a) and (c) is the specific heat of an ideal Bose gas with $(N_{\rm B}^x,N_{\rm B}^y,N_{\rm B}^z)=(N/6,N/6,N/6)$ and with $(N/2,0,0)$, respectively (Note that these two cases give the same value of the specific heat at $T_{\rm c}$.)}
\label{fig5}
\end{figure}
\par
To understand the above-mentioned behavior of $C_V(T)$, Fig. \ref{fig5} compares $C_V(T)$ with the molecular numbers $N_{\rm B}^{i=x,y,z}$ in Eq. (\ref{eq11b}) in the strong-coupling case ($(k_{\rm F}v_x)^{-1}=15$). When $\delta v^{-1}=0$, Fig. \ref{fig5}(a) shows that $N_{\rm B}^x=N_{\rm B}^y=N_{\rm B}^z\simeq N/6$ at low temperatures. As a result, $C_V(T)$ below the dip temperature ($\equiv T_{\rm dip}\simeq 0.2T_{\rm F}$) is well described by the specific heat $C_V^{\rm B}(T)$ in an ideal Bose gas with three kinds of $N/6$ $p$-wave molecules. The enhancement of $C_V(T\simeq T_{\rm c})$ in this case is simply due to the well-known temperature dependence of the specific heat in an ideal Bose gas neat the BEC transition temperature. The deviation from $C_V^{\rm B}(T)$ above $T\simeq T_{\rm dip}$ reflects the onset of thermal dissociation of $p$-wave molecules.
\par
When $(k_{\rm F}^3\delta v)^{-1}=2$, Fig. \ref{fig5}(b) shows that, although the overall behavior of $C_V$ looks similar to the case in panel (a), the origin of the increase of $C_V(T)$ below $T_{\rm dip}$ is due to the remarkable increase (decrease) of $N_{\rm B}^x$ ($N_{\rm B}^{i=y,z}$) with decreasing the temperature, contributing to $C_V^{(2)}$ in Eq. (\ref{eq17b}). Although $N_{\rm B}^x$ does not reach $N/2$ at $T_{\rm c}$ in Fig. \ref{fig5}(b), such a situation is realized near $T_{\rm c}$ in the case of Fig. \ref{fig5}(c). In the latter case, when $N_{\rm B}^{i=y,z}/(N/2)\ll 1$, their weak temperature dependence only gives small values of $C_V^{(2)}$. As a result, $C_V(T)$ again decreases with decreasing the temperature, leading to the hump structure around $T/T_{\rm F}\simeq 0.5~(\equiv T_{\rm hump})$ in Fig. \ref{fig5}(c). Since the region near $T_{\rm c}$ is well described by an ideal Bose gas with $N/2$ $p_x$-wave molecules, $C_V(T)$ also exhibits a dip structure at $T/T_{\rm F}\simeq 0.2$, reflecting the behavior of $C_V^{\rm B}$ in the ideal Bose gas near $T_{\rm c}$, as shown in Fig. \ref{fig5}(c).
\par
\begin{figure}
\centerline{\includegraphics[width=8cm]{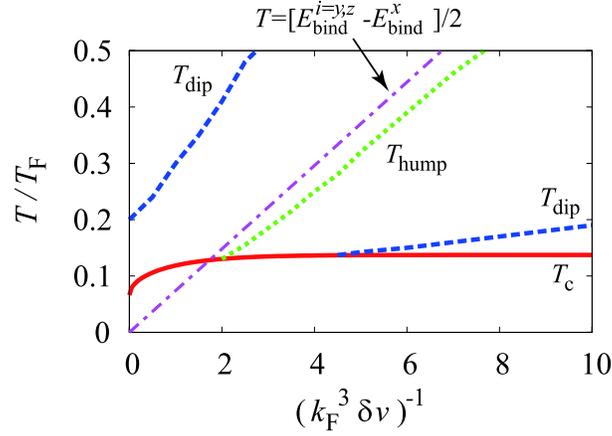}}
\caption{(Color online) Characteristic temperatures $T_{\rm dip}$, and $T_{\rm hump}$ in the strong-coupling regime. We take $(k_{\rm F}^3v_x)^{-1}=15$. For the definitions of these temperatures, see the text. The dashed-dotted line shows half the energy difference $\Delta E=E_{\rm bind}^{i=y,z}-E_{\rm bind}^x$.}
\label{fig6}
\end{figure}
\par
Figure \ref{fig6} summarizes two dip temperatures $T_{\rm dip}$, as well as the hump temperature $T_{\rm hump}$, determined from the temperature dependence of the specific heat $C_V(T)$ in the strong-coupling regime ($(k_{\rm F}^3v_x)^{-1}=15\gg 1$). In this case, the molecular binding energy is evaluated as $E_{\rm bind}^x=2\varepsilon_{\rm F}$ (where $\varepsilon_{\rm F}$ is the Fermi energy). Thus, the system is considered to be dominated by tightly bound $p$-wave molecules in the temperature region shown in Fig. \ref{fig6}. While the three kinds of $p_i$-wave molecules ($i=x,y,z$) are nearly equally populated above $T_{\rm dip}$, the $p_x$-wave component gradually becomes dominant below $T_{\rm dip}$. Below the lower dip temperature near $T_{\rm c}$, the system may be viewed as an ideal Bose gas of $N/2$ $p_x$-wave wave molecules. The characteristic energy scale in this continuous change from a ``three-component" Bose gas to a ``one-component" Bose gas is given by the energy difference $\Delta E=E_{\rm bind}^{i=y,z}-E_{\rm bind}^x$ between the higher and lower molecular bound states. Indeed, Fig. \ref{fig6} shows
\begin{equation}
T_{\rm hump}\simeq {\Delta E \over 2}.
\label{eqq}
\end{equation}
 This result is consistent with the peak temperature of the specific heat in a simple two-level system with the energy difference $\Delta E~(\gg T)$\cite{note}. Here we note that, since within the NSR theory the molecular-molecular interaction is not taken into account, this population imbalance among the three components of the $p$-wave molecular boson is not induced by the boson-boson interaction, but by thermally transferring from the high-energy $p_y$ and $p_z$-wave states to the low-energy $p_x$-wave states as decreasing temperature, due to the binding energy difference $\Delta E$. 
\par
\begin{figure}
\centerline{\includegraphics[width=8cm]{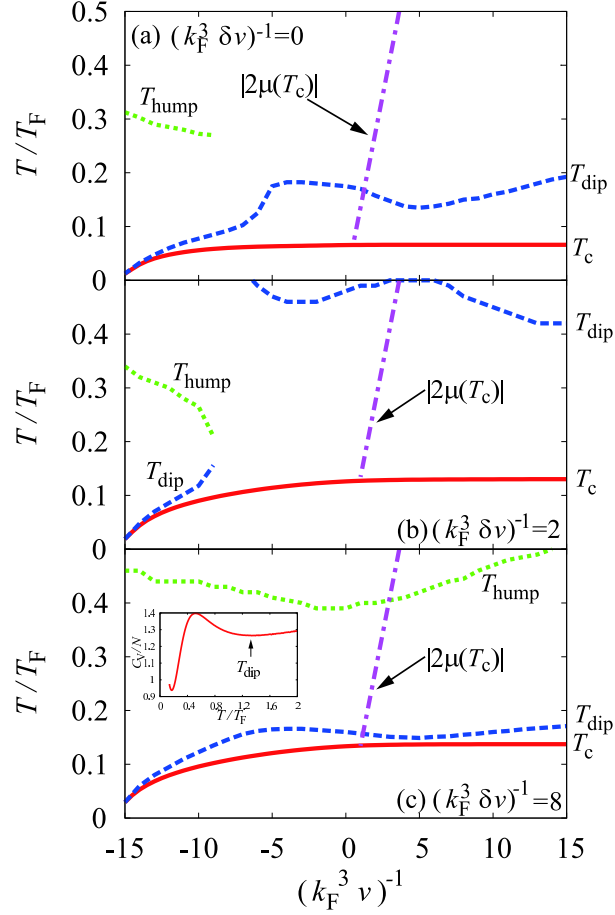}}
\caption{(Color online) Characteristic temperatures $T_{\rm dip}$ and $T_{\rm hump}$ drawn in the phase diagram of a uniaxially anisotropic $p$-wave interacting Fermi gas, in terms of the $p$-wave interaction strength and the temperature. We also plot $|2\mu(T_{\rm c})|$ in the strong-coupling regime when $\mu(T_{\rm c})<0$, which physically gives a characteristic temperature below which two-body bound molecules are gradually formed, overwhelming thermal dissociation. The inset in panel (c) shows $C_V(T)$ when $(k_{\rm F}^3 v_x)^{-1}=15$ and $(k_{\rm F}^3\delta v)^{-1}=8$, to show the magnitude of higher dip temperature.}
\label{fig7}
\end{figure}
\par
Although $T_{\rm dip}$ and $T_{\rm hump}$ are merely crossover temperatures without being accompanied by any phase transition, it is still interesting to plot them in the phase diagram of a $p$-wave interaction Fermi gas (Fig. \ref{fig7}), to understand normal state properties of a $p$-wave interacting Fermi gas, especially in the strong-coupling regime. When $\mu/\varepsilon_{\rm F}\ll -1$ in this regime, the $T_{\rm c}$ equation (\ref{eq13}) is reduced to
\begin{equation}
1=U_x\sum_{\bm p}
{(\gamma_{\bm p}^x)^2 \over 2\varepsilon_{\bm p}+2|\mu(T_{\rm c})|},
\label{eq19}
\end{equation}
which is just the same form as the equation for a two-body bound state with the binding energy $E_{\rm bind}^x=-2|\mu(T_{\rm c})|$. Thus, the line ``$2|\mu(T_{\rm c})|$" in Fig. \ref{fig7} physically gives a characteristic temperature around which two-body $p$-wave bound molecules appear, overwhelming thermal dissociation\cite{Tsuchiya1}.
\par
When $\delta v^{-1}=0$, Fig. \ref{fig7}(a) shows that two-body bound molecules are gradually formed with decreasing the temperature in the right side of the $2|\mu(T_{\rm c})|$-line. Below $T_{\rm dip}$, the system becomes close to an ideal Bose gas with three kinds of $N/6$ $p$-wave molecules, in the sense that $C_V(T)$ agrees well with the specific heat in this ideal Bose gas.
\par
The dip temperature in the strong-coupling side increases to $T_{\rm dip}=0.4\sim 0.5T_{\rm F}$, when $(k_F^3\delta v)^{-1}=2$ (see Fig. \ref{fig7}(b)). In this case, while the three $p_i$-wave molecules ($i=x,y,z$) are nearly equally populated in the right side of the dashed-dotted line above $T_{\rm dip}$, the $p_x$-wave component gradually become dominant below $T_{\rm dip}$. However, the molecular transitions to the $p_x$-wave state from the $p_y$- and $p_z$-wave state do not complete even at $T_{\rm c}$, so that the region where $C_V(T)$ can be described by the specific heat in an ideal Bose gas does not exist in this case. Such a region is obtained when one further increases the anisotropy parameter $\delta v^{-1}$, as shown in Fig. \ref{fig7}(c). In this figure, the hump temperature $T_{\rm hump}\sim \Delta E=[E_{\rm bind}^{i=y,z}-E_{\rm bind}^x]/2$ appears below the dip temperature ($T_{\rm dip}\simeq 1.2T_{\rm F}$, see the inset in this panel). In addition, we obtain the second dip temperature near $T_{\rm c}$, below which the system may be viewed as an ideal Bose gas with $N/2$ $p_x$-wave molecules.
\par
As discussed in our previous papers\cite{Inotani4,Inotani5}, the low temperature region ($T\lesssim T_{\rm dip}$) in the weak-coupling side ($(k_F^3v_x)^{-1}<0$) in Fig. \ref{fig7}(a) is dominated by $p$-wave pairing fluctuations, leading to the enhancement of $C_V(T)$. On the other hand, the hump temperature $T_{\rm hump}=0.2\sim 0.3T_{\rm F}$ in the weak-coupling side in Fig. \ref{fig7}(a) originates from the enhancement of $C_V$ by anomalous particle-particle scatterings into $p$-wave molecular excitations\cite{Inotani4}. When $(k_{\rm F}^3 \delta v)^{-1}=2$, Fig. \ref{fig7}(b) shows that, while this hump temperature remains, the dip temperature exhibits a discontinuity at $(k_{\rm F}^3v_x)^{-1}\sim -8$ within the numerical accuracy. As seen in Fig. \ref{fig4}(a), because the uniaxial anisotropy of the $p$-wave interaction is not crucial in the weak-coupling regime, $T_{\rm dip}$ in the region $(k_{\rm F}^3v_x)^{-1}\lesssim-8$ in Fig. \ref{fig7}(b) is considered to have the same physical meaning as in the case of panel (a).  On the other hand, $T_{\rm dip}$ in the region $(k_{\rm F}^3v_x)^{-1}\gesim -8$ smoothly connects to $T_{\rm dip}$ obtained in the strong-coupling side. Thus, although a two-body bound state is absent in the weak-coupling side, $T_{\rm dip}$ around $-8\lesssim (k_{\rm F}^3v_x)^{-1}\lesssim0$ in Fig. \ref{fig7}(b), as well as $T_{\rm hump}$ in the weak-coupling side of Fig. \ref{fig7}(c), may be associated with an imbalance effect among three $p_i$-wave Cooper channels $(i=x,y,z)$. In this regard, we point out that  the spectrum of the analytic continued particle-particle scattering matrix in Eq. (\ref{eq.Gamma}) is known to still have a sharp peak along the molecular kinetic energy in Eq. (\ref{eq16b}) with $E_{\rm bind}^i=0$, even in the weak-coupling regime\cite{Inotani}. Although we need further analyses to clarify background physics of $T_{\rm dip}$ and $T_{\rm hump}$ around $T=0.4\sim 0.5T_{\rm F}$ in the weak-coupling side of Figs. \ref{fig7}(b) and (c), this fact makes us expect that they may have similar physical meanings to those in the strong-coupling side.
\par
When $(k_{\rm F}^3 \delta v)^{-1}=8$ shown in Fig. \ref{fig7}(c), the system is dominated by $p_x$-wave pairing fluctuations near $T_{\rm c}$, so that $T_{\rm dip}$ in the weak-coupling side near $T_{\rm c}$ may be safely regarded as the characteristic temperature below which $p_x$-wave pairing fluctuations enhance $C_V(T)$.
\par
\par
\section{Summary}
\par
To summarize, we have discussed normal state properties of an ultracold Fermi gas and effects of uniaxial anisotropy ($U_x>U_y=U_z)$ of a $p$-wave pairing interaction. In particular, we have dealt with the specific heat $C_V(T)$ at constant volume, as an example of observable thermodynamic quantity in this system. Including $p$-wave pairing fluctuations within the framework of the strong-coupling theory developed by Nozi\`eres and Schmitt-Rink, we have clarified how $C_V(T)$ is affected by the uniaxial anisotropy of the $p$-wave interaction, from the weak- to strong-coupling regime. 
\par
At $T_{\rm c}$, we showed that, while the uniaxial anisotropy is not crucial for $C_V$ in the weak-coupling regime, it is largely enhanced in the strong-coupling regime when $(k_{\rm F}\delta v)^{-1}\simeq 1$. This is because $T_{\rm c}$ in this case is comparable to the energy difference between the binding energies $E_{\rm bind}^{i=y,z}$ in the $p_y$- and $p_z$-wave channels and the binding energy $E_{\rm bind}^x$ ($<E_{\rm bind}^{i=y,z}$) in the $p_x$-wave channel, so that molecular transitions from the former two states to the latter with decreasing the temperature lead to the enhancement of $C_V(T_{\rm c})$. Such an effect is absent when $\delta v^{-1}=0$ and $(k_{\rm F}^3\delta v)^{-1}\gg 1$. In these cases, $C_V(T_{\rm c})$ in the strong-coupling regime are simply described by the specific heat in an ideal Bose gas mixture with three kinds of $N/6$ $p$-wave molecules and a one-component ideal Bose gas consisting of $N/2$ $p_x$-wave molecules, respectively (where $N$ is the number of Fermi atoms).
\par
We also clarified that the above-mentioned molecular transition also affects the behavior of $C_V(T)$ above $T_{\rm c}$, especially in the strong-coupling side ($(k_{\rm F}^3v_x)^{-1}\ge 0$). In this regime, with decreasing the temperature, we showed that $C_V(T)$ exhibits a dip structure at the temperature $T_{\rm dip}$, around which the population imbalance starts to occur among $p_i$-wave molecules ($i=x,y,z$) (that have already been formed at $T\sim 2|\mu|>T_{\rm dip}$). With further decreasing the temperature, $C_V(T)$ exhibits a hump structure at $T\sim [E_{\rm bind}^{i=y,z}-E_{\rm bind}^x]/2$. When most molecules occupy the lowest $p_x$-wave state, $C_V(T)$ again shows a dip structure, below which the temperature dependence is well described by the specific heat in an ideal Bose gas with $N/2$ $p_x$-wave molecules. Our results indicate that $C_V(T)$ is a useful thermodynamic quantity to see the molecular character in the strong-coupling regime of a $p$-wave interacting Fermi gas, when the interaction possesses a uniaxial anisotropy. 
\par
The dip temperature $T_{\rm dip}$ (which gives the onset of the population imbalance among the three $p$-wave molecules), as well as the hump temperature $T_{\rm hump}$ (which is comparable to half the energy difference between the binding  energies $\Delta E=[E_{\rm bind}^{i=y,z}-E_{\rm bind}^x]/2$), obtained in the strong-coupling side continue to exist in the weak-coupling side ($(k_{\rm F}v_x)^{-1}\le 0$). In this regime, a two-body bound molecule no longer exists, so that it is still unclear whether or not the physical interpretations for $T_{\rm dip}$ and $T_{\rm hump}$ obtained in the strong-coupling regime are also valid for the weak-coupling case, which remains as our future problem. However, the known fact that the $p$-wave pair correlation function still has a sharp spectral peak along the molecular dispersion even in the weak-coupling regime\cite{Inotani,Inotani2} implies validity of the interpretations obtained in the strong-coupling side to the weak-coupling side to some extent. 
\par  
In the weak-coupling regime, we also obtained another dip temperature near $T_{\rm c}$, below which $p$-wave pairing fluctuations become strong, leading to the enhancement of $C_V$.
\par
Since the discovery of the splitting of a $p$-wave Feshbach resonance in a $^{40}$K Fermi gas\cite{Ticknor}, the importance of the associated uniaxially anisotropic $p$-wave pairing interaction has mainly been discussed in the context of multi-superfluid phase below $T_{\rm c}$. Of course, the realization of a $p$-wave superfluid state is the most important issue in the study of $p$-wave interacting Fermi gases. However, at present, this challenge is facing serious difficulties, such as three-body loss\cite{Castin,Gurarie3}, as well as dipolar relaxation\cite{Gaebler2}. Thus, as an alternative approach to this non-$s$-wave system, it would also be a useful strategy to start from the study of normal state properties above $T_{\rm c}$. Since the specific heat has recently become observable in cold Fermi gas physics, our results would contribute to the further development of research on $p$-wave interacting Fermi gases, when the interaction is uniaxially anisotropic. 
\par
\begin{acknowledgments}
YO thanks T. Mukaiyama for useful discussions on the stability of a $p$-wave interacting Fermi gas. This work was supported by KiPAS project in Keio University. DI was supported by Grant-in-aid for Scientific Research from JSPS in Japan (No.JP16K17773). YO was supported by Grant-in-aid for Scientific Research from MEXT and JSPS in Japan (No.JP15H00840, No.JP15K00178, No.JP16K05503). 
\end{acknowledgments}
\par


\begin{thebibliography}{99}
\bibitem{note} We briefly note that the discussion here is somehow different from that around Eq. (\ref{eq17c}). While the former discussion considers the case above $T_{\rm c}$ for a given value of the anisotropy parameter (leading to $T_{\rm hump}\simeq \Delta E/2$), the latter treats the situation along the $T_{\rm c}$-line in Fig. \ref{fig7} (leading to $T_{\rm c}\simeq \Delta E$ around the hump structure seen in Fig. \ref{fig2}(b)). However, apart from the difference by the factor $1/2$, the background physics that molecular excitations play an essential role is the same between the two cases.
\end{thebibliography}
\end{document}